\documentclass{article}
 \usepackage{amssymb}
 \usepackage{epsfig}
 \newtheorem{Lemma}{Lemma}

 \newtheorem{Theorem}[Lemma]{Theorem}

 \newcommand{\Ex}{{\mathbb{E}}}
 
 \renewcommand{\Pr}{{\mathbb{P}}}

 \newcommand{\EE}{\mbox{${\mathcal E}$}}

 \newcommand{\bx}{{\mathbf x}}

 \newcommand{\sfrac}[2]{{\textstyle\frac{#1}{#2}}}

 \newcommand{\cost}{\mathrm{cost}}

 \newcommand{\ave}{\mathrm{ave}}

\newcommand{\Gil}{\mathrm{Gil}}
 \begin{document}
 \title{Optimal spatial transportation networks where link-costs are sublinear in link-capacity}
 \author{D J Aldous\thanks{Research supported by
 N.S.F Grant DMS-0704159}\\
Department of Statistics\\
 367 Evans Hall \#\  3860\\
 U.C. Berkeley CA 94720\\
aldous@stat.berkeley.edu
 \\ www.stat.berkeley.edu/users/aldous}

  \maketitle

\begin{abstract}
Consider designing a transportation network on $n$ vertices in the plane, with traffic demand uniform over all source-destination pairs. 
Suppose the cost of a link of length $\ell$ and capacity $c$ scales as $\ell c^\beta$ for fixed $0<\beta<1$.  
Under appropriate standardization, 
the cost of the minimum cost {\em Gilbert network} grows essentially as 
$n^{\alpha(\beta)}$, where 
$\alpha(\beta) = 1 - \frac{\beta}{2}$ on $0 < \beta \leq \frac{1}{2}$ and 
$\alpha(\beta) = \frac{1}{2} + \frac{\beta}{2}$ on $\frac{1}{2} \leq \beta < 1$.
This quantity is an upper bound in the worst case (of vertex positions), and a lower bound under 
mild regularity assumptions.  
Essentially the same bounds hold if we constrain the network to be \emph{efficient} in the sense that average route-length is only 
$1 + o(1)$ times average straight line length. 
The transition at $\beta = \frac{1}{2}$ corresponds to the dominant cost contribution changing from short links to long links.  
The upper bounds arise in the following type of hierarchical networks, 
which are therefore optimal in an order of magnitude sense. 
On the large scale, use a sparse Poisson line process to provide 
long-range links.
On the medium scale, use hierachical routing on the square lattice.
On the small scale, link vertices directly to medium-grid points.
We discuss one of many possible variant models, in which links also have a designed maximum speed 
$s$ and the cost becomes $\ell c^\beta s^\gamma$.
\end{abstract}


 \vspace{0.4in}


 \section{Introduction}
 \label{sec-INT}
 To design a transportation network linking specified points (visualized as cities) in the plane, 
one might specify a cost functional
and a benefit functional on all possible networks,
and then consider networks which are optimal in the sense
of minimizing cost for a given level of benefit.
This paper addresses one particular choice of functionals,
but our broader purpose 
(see section \ref{sec-methodology}) 
is to draw the attention of statistical 
physicists to this class of problem.

We study a simple model involving the
 ``economy of scale" idea
 \begin{quote}
 One link of length $\ell$ and capacity $2c$ is less than twice as expensive as 
 two links of length $\ell$ and capacity $c$.
 \end{quote}
We capture this idea by specifying that the
 cost of  a link of length $\ell$ and capacity $c$ scales as $\ell c^\beta$ 
 for some $0 < \beta < 1$.
 In the real world, network designers do not know in advance what traffic demand will be.  
 We simplify by assuming that traffic demand is known (and uniform over all source-destination pairs) 
 and routes are controlled, so that the volume $f(e)$ of flow across an edge (link) $e$ can be determined by the designers, and the corresponding link-capacity built.  
 (Visualize links as roads, and flow-volume $f(e)$  as ``number of vehicles per hour".  
 We are ignoring stochastic fluctuations in traffic). Thus our cost structure is
 \begin{equation}
   \mbox{cost of network } = \sum_e \ell(e) f^\beta(e) 
   \label{cost-def} 
   \end{equation}
where $\ell(e) = $ length of link $e$.

To define the model carefully, write 
$\bx^n = \{x_1,x_2,\ldots,x_n\}$
for a configuration of $n$ vertices in the square 
$[0,n^{1/2}]^2$ 
of area $n$.  
So $x_i$ is the position of vertex $i$.
 Create a connected {\em network}  $G(\bx^n)$ by adding links:
links are line-segments with their natural Euclidean lengths,
and links may meet at places not in the given vertex-set $\bx^n$. 
To make the distinction clear let us refer to the given $n$ vertices as {\em cities} and any meeting places 
(which depend on our choice of network) as {\em junctions}.
Between each source-destination pair $(i,j)$ of cities, flow of volume $n^{-3/2}$ 
(this scaling is explained below) is routed through the network.  
Define $\cost(G(\bx^n))$, the cost of the network, via (\ref{cost-def}).  
This setting specializes a setting considered by Gilbert \cite{gilbert-MCCN}, and we call the minimum-cost network the {\em Gilbert network} $\Gil(\bx^n)$.   
See \cite{thomas-weng} for general properties of, and heuristic algorithms for, Gilbert networks over deterministic points.

Gilbert networks may be optimal from a network {\em operator} viewpoint, but what about a network {\em user}?
Write $\ell(x_i,x_j)$ for route-length, and $|x_j - x_i|$ for straight-line distance, between cities $i$ and $j$. For a typical configuration, the average 
distance 
 $\ave_{i,j} |x_j - x_i|$ will be order $n^{1/2}$. 
 The kind of ``benefit to users" we have in  mind is that the network provides routes almost as short as possible.
 So we call the sequence of networks $(G(\bx^n))$ 
 \emph{modestly efficient} 
 if
  \begin{equation}
 \ave_{i,j} (\ell(x_i,x_j) - |x_j - x_i|) = o(n^{1/2}) . \label{modest}
 \end{equation} 
 The name reflects the remarkable fact \cite{me116} that there exist \emph{extremely efficient} networks for which  this average is  $O(\log n)$ while their length is only $1 + o(1)$ times the minimum length of any connected network;  such results pay no attention to flow-volumes or capacities, and so constitute the $\beta = 0$ case of the present model.
The problem we address in this paper is:
\begin{quote}
   given the sequence $(\bx^n)$, how small can we make $\cost(G(\bx^n))$ subject to the 
  \emph{modestly efficient} constraint (\ref{modest})?
\end{quote}
In the $\beta = 0$ case just mentioned, we can make
$\cost(G(\bx^n))$ 
be asymptotically the length of the Steiner tree (minimum length connected network)
on $\bx^n$, which is well known to be $O(n)$ in the worst case and in the typical case.
Recall that 
$a_n = O(b_n)$ means that 
$a_n/b_n$ is bounded as $n \to \infty$.
It is often convenient to write the converse relationship $b_n = O(a_n)$ as 
$a_n = \Omega(b_n)$; 
if both $a_n = O(b_n)$ and $a_n = \Omega(b_n)$ then we write 
$a_n = \Theta(b_n)$.

In the case $\beta = 1$ there is no 
``economy of scale" and so the minimum-cost network is just the complete graph, 
that is a direct link between each pair of cities.
The associated cost is
\[ \sum_i \sum_j n^{-3/2} |x_i - x_j|
= n \times \frac{\ave_{i,j} |x_i - x_j|}{n^{1/2}} \]
which is $O(n)$ in the worst case and in the typical case. 

Recall that the Gilbert network $\Gil(\bx^n)$ is the minimum-cost network when there is no extra ``modestly efficient" constraint.  Theorem \ref{T1} shows that imposing the ``modestly efficient" constraint makes little difference in an order of magnitude sense: in either case the optimal cost grows roughly as order $n^{\alpha(\beta)}$.
\begin{Theorem}
\label{T1}
Fix $0 < \beta < 1$.
Define 
\begin{eqnarray*}
\alpha(\beta) &=& 1 - \sfrac{\beta}{2}, \quad 0 < \beta \leq \sfrac{1}{2}\\
&=& \sfrac{1+\beta}{2}, \quad \sfrac{1}{2} \leq \beta < 1.
\end{eqnarray*}
Let $\bx^n$
be a configuration of $n$ cities in the square 
$[0,n^{1/2}]^2$.
\\ (a)  {\bf Case $ 0 < \beta \leq \sfrac{1}{2}$.} There exist modestly efficient networks for which 
$\cost(G(\bx^n)) = O(n^{\alpha(\beta)})$ (except for $\beta = 1/2$ the bound is $O(n^{3/4} \log n)$).  
Under the technical assumption (\ref{LB1}) there do \emph{not} exist connected networks  for which 
$\cost(G(\bx^n)) = o(n^{\alpha(\beta)})$.  
So under (\ref{LB1}) we have $\cost(\Gil(\bx^n)) = \Theta(n^{\alpha(\beta)})$ for $\beta < \sfrac{1}{2}$.
\\ (b) {\bf Case $ \sfrac{1}{2} < \beta < 1$.}  
Here $\cost(\Gil(\bx^n)) = O(n^{\alpha(\beta)})$.  
Given  $\omega_n \to \infty$ arbitrarily slowly,
there exist modestly efficient networks for which
$\cost(G(\bx^n)) = O(\omega_n n^{\alpha(\beta)})$. 
Under the technical assumption  (\ref{LB2}), $\cost(\Gil(\bx^n)) = \Theta(n^{\alpha(\beta)})$, but 
there do \emph{not} exist modestly efficient networks for which 
$\cost(G(\bx^n)) = O(n^{\alpha(\beta)})$.
\end{Theorem}
Our discussion above of the cases $\beta = 0$ and $\beta = 1$ 
implies corresponding results in these cases with 
$\alpha(0) = 1$ and $\alpha(1) = 1$.

The transition at $\beta = \frac{1}{2}$ corresponds to the dominant cost contribution changing from short links to long links,  
as we will explain in section \ref{sec-explain}.
The technical regularity assumptions that we need to impose to obtain lower bounds 
reflect this transition: 
for $\beta < 1/2$ we need to assume that nearest-neighbor distances are not atypically small,
whereas for $\beta > 1/2$ we assume a large-scale equidistribution of the city configuration.  
(We defer statements of these assumptions until the place they are actually used in the proof, to avoid interrupting the conceptual discussion here.)
We show (section \ref{sec-upper}) that 
the upper bounds arise in the following type of hierarchical networks, 
which are therefore optimal in an order of magnitude sense. 
On the large scale, use a sparse Poisson line process to provide 
long-range links.
On the medium scale, use hierachical routing on the square lattice.
On the small scale, link cities directly to medium-grid points.
It is perhaps counter-intuitive that one can use the same network for the whole range of $\beta$; 
the point is that only the medium-small scale structure really matters 
for $\beta < 1/2$ and only the large scale structure really matters for 
$\beta > 1/2$.
Our arguments implicitly imply some weak properties of the 
{\em exactly} optimal networks.
Undestanding in detail the structure of the Gilbert network 
(or the asymptotically optimal modestly efficient network)
over random points 
in the critical case $\beta = 1/2$
is a challenging problem, interesting because one expects the network to have some
{\em scale-free} stucture, in the (correct) sense of invariance under spatial and flow-volume rescaling.

One can imagine many variant models in which extra structure is incorporated. 
In section \ref{sec-speed} we briefly discuss the case where links have designed speed $s$ and where the cost of a link becomes $\ell c^\beta s^\gamma$; in this case 
an analog of Theorem \ref{T1} remains true.

\subsection{Optimal spatial network design methodology}
\label{sec-methodology}
This paper contributes to a general program concerning networks linking points in the plane:
\begin{quote}
for mathematically simple cost/benefit functionals, study the properties 
(geometry, cost and benefit values)
of optimal networks as the number $n$ of points tends to infinity. 
\end{quote}
Network design problems arise in many applied fields, 
but serious real-world modelling leads to more complicated functionals tuned to specific applications than we have in mind.
As complementary work, 
\cite{me116} gives a detailed treatment of
the {\em extremely efficient} networks mentioned above 
that minimize average route length subject to total network length;
and \cite{me118} analyzes a model 
(for e.g. passenger air travel or package delivery) 
where there is a substantial cost to transfer from one link to another.
In the latter model,
theory predicts that hub-and-spoke networks (as seen in the real world) are near-optimal 
and that, constraining the average number of transfers to be say $2$, the length of the shortest possible network scales as $n^{13/10}$.

The methodological feature we want to emphasize concerns models for the position of $n$ cities 
(assumed for simplicity in a square of area $n$).
In each problem we have studied one gets the same order of magnitude for optimal network cost 
for worst-case positions as one gets for arbitrary positions (under mild assumptions) and in particular 
the same as for random positions or for regular (e.g. lattice) positions.

The bulk of statistical physics literature on spatial networks
(surveyed in 
\cite{hayashi-2006-47})
analyzes networks built according to some specific probability model 
which combines ingredients such as 
\\ (a) 
geometric random graphs 
(link probability depends on inter-vertex distance); 
\\ (b) 
proportional attachment probabilities for arriving vertices;
\\ (c) prescribed power law distribution of lattice vertex degrees;
\\ (d) networks based on recursive partitioning of space. \\
This theoretical literature makes passing reference to optimality, but
we have not seen analytic results demonstrating optimality over 
{\em all} possible networks in the spatial context 
(see \cite{donetti06} for non-spatial results, and \cite{bejan-book,barth06} for assumptions under which optimal networks are trees).
For interesting empirical work see
\cite{gastner-newman}.

 Our scaling conventions (a square of area $n$; flow-volume $n^{-3/2}$ between each source-destination pair) may seem arbitrary, but are chosen to fit the following standardizations: \\
 (i) cities have density $1$ per unit area;\\
 (ii) flow volume across unit area is order $1$.

\section{The construction}
\label{sec-upper}
A network satisfying the requirements of Theorem \ref{T1}
will be constructed in section \ref{sec-tc} using
mathematical ingredients described in sections \ref{sec-hier} and \ref{sec-PLP}.   
Figure 1 illustrates the construction.

\setlength{\unitlength}{0.09in}
\begin{picture}(42,34)(0,-2)
\put(0,5){\line(1,0){16}}
\put(0,5.03){\line(1,0){16}}
\put(0,5.06){\line(1,0){16}}
\put(0,5.09){\line(1,0){16}}
\put(0,6){\line(1,0){16}}
\put(0,7){\line(1,0){16}}
\put(0,7.03){\line(1,0){16}}
\put(0,8){\line(1,0){16}}
\put(0,9){\line(1,0){16}}
\put(0,9.03){\line(1,0){16}}
\put(0,9.06){\line(1,0){16}}
\put(0,10){\line(1,0){16}}
\put(0,11){\line(1,0){16}}
\put(0,11.03){\line(1,0){16}}
\put(0,12){\line(1,0){16}}
\put(0,13){\line(1,0){16}}
\put(0,13.03){\line(1,0){16}}
\put(0,13.06){\line(1,0){16}}
\put(0,13.09){\line(1,0){16}}
\put(0,14){\line(1,0){16}}
\put(0,15){\line(1,0){16}}
\put(0,15.03){\line(1,0){16}}
\put(0,16){\line(1,0){16}}
\put(0,17){\line(1,0){16}}
\put(0,17.03){\line(1,0){16}}
\put(0,17.06){\line(1,0){16}}
\put(0,18){\line(1,0){16}}
\put(0,19){\line(1,0){16}}
\put(0,19.03){\line(1,0){16}}
\put(0,20){\line(1,0){16}}

\put(0,5){\line(0,1){16}}
\put(1,5){\line(0,1){16}}
\put(2,5){\line(0,1){16}}
\put(2.03,5){\line(0,1){16}}
\put(3,5){\line(0,1){16}}
\put(4,5){\line(0,1){16}}
\put(4.03,5){\line(0,1){16}}
\put(4.06,5){\line(0,1){16}}
\put(5,5){\line(0,1){16}}
\put(6,5){\line(0,1){16}}
\put(6.03,5){\line(0,1){16}}
\put(7,5){\line(0,1){16}}
\put(8,5){\line(0,1){16}}
\put(8.03,5){\line(0,1){16}}
\put(8.06,5){\line(0,1){16}}
\put(8.09,5){\line(0,1){16}}
\put(9,5){\line(0,1){16}}
\put(10,5){\line(0,1){16}}
\put(10.03,5){\line(0,1){16}}
\put(11,5){\line(0,1){16}}
\put(12,5){\line(0,1){16}}
\put(12.03,5){\line(0,1){16}}
\put(12.06,5){\line(0,1){16}}
\put(13,5){\line(0,1){16}}
\put(14,5){\line(0,1){16}}
\put(14.03,5){\line(0,1){16}}
\put(15,5){\line(0,1){16}}
\put(-1,13){\vector(0,1){1}}
\put(-3,13.2){$s_n$}
\put(2,23){$2^{M_n}$ small cells}
\put(0,22.4){\vector(1,0){16}}
\put(17,5){\vector(0,1){16}}
\put(17.6,10.2){$\sigma_n$}
\put(12,14.7){$\triangleleft$}
\put(11.5,13.6){$\triangledown$}
\put(10,12.7){$\triangleleft$}
\put(7.4,9){$\triangledown$}
\put(4,4.7){$\triangleleft$}
\put(25,0){\line(1,0){25}}
\put(25,5){\line(1,0){25}}
\put(25,10){\line(1,0){25}}
\put(25,15){\line(1,0){25}}
\put(25,20){\line(1,0){25}}
\put(25,0){\line(0,1){25}}
\put(30,0){\line(0,1){25}}
\put(35,0){\line(0,1){25}}
\put(40,0){\line(0,1){25}}
\put(45,0){\line(0,1){25}}
\put(25,25){\line(1,0){25}}
\put(50,0){\line(0,1){25}}
\put(23,15){\vector(0,1){5}}
\put(21,17.5){$\sigma_n$}
\put(33,27){$\theta_n$ large cells}
\put(25,26.4){\vector(1,0){25}}
\put(52,0){\vector(0,1){25}}
\put(52.6,13){$n^{1/2}$}
\put(25,12){\line(1,1){13}}
\put(28,25){\line(1,-1){22}}
\put(26,0){\line(2,1){24}}
\put(25,21){\line(2,-1){25}}
\put(44,25){\line(1,-2){6}}
\put(29,0){\line(1,2){12.5}}
\put(25,16){\line(1,-3){5.33}}
\put(42,0){\line(1,3){8}}
\put(25,6){\line(3,-1){18}}
\put(25,13){\line(3,1){25}}
\end{picture}

\noindent
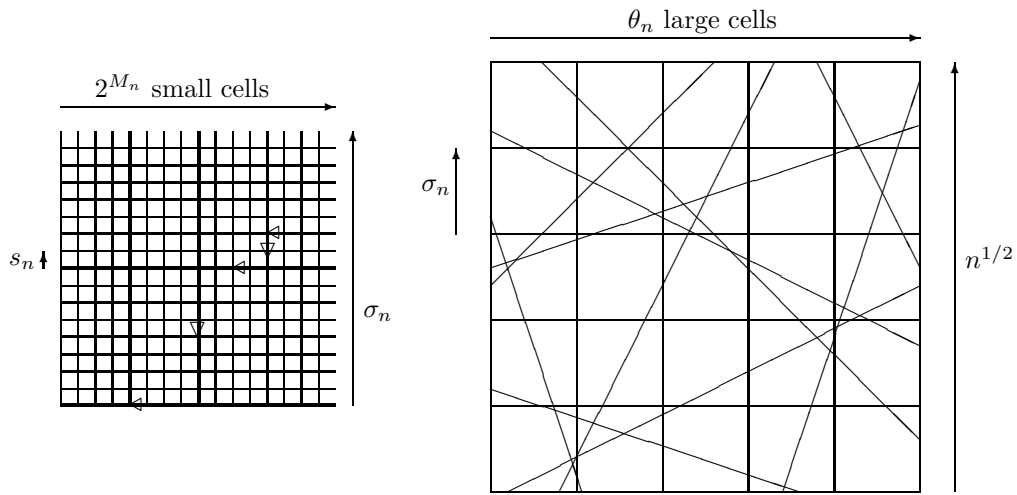
{\bf Figure 1.   Ingredients of the construction.}
{\small 
Left: 
the hierarchical routing lattice, with higher-type edges indicated by thicker lines,
and a typical route shown.
Right: the large-scale grid and the Poisson line process.
}

\subsection{Hierarchical routing on the square lattice}
\label{sec-hier}
Fix $M$ and consider the square grid on vertices 
$\{0,1,2,\ldots,2^M -1\}^2$.
Declare lines (and their edges) to be of some type 
$0,1,2,\ldots,M$
according to the rule:

the horizontal lines 
$\{(x,y): y = (2j-1)2^m\}, \quad j = 1,2,\ldots  $
are type $m$

the boundary line $\{(x,0)\}$ is type $M$;

\noindent
and similarly for vertical lines.
For each vertex $(x,y)$, define a route from $(x,y)$ to $(0,0)$ using only
downward and leftward edges as follows.
First choose the edge at $(x,y)$ of higher type (breaking ties arbitrarily).
Then repeat the rule
\begin{quote}
Follow the current edge until it crosses an edge of strictly higher type, then 
transfer to that edge 
\end{quote}
until reaching $(0,0)$.
See Figure 1, left side.

It is elementary to verify
\begin{Lemma}
\label{LmM}
For each $0 \leq m \leq M$, 
the number of type-$m$ edges traversed by the route is at most 
$2^{m+1}$.
\end{Lemma}

\subsection{The Poisson line process}
\label{sec-PLP}
A line in the plane may be parametrized by the point $z$ on the line which is closest to the origin 
(so the line segment from the origin to $z$ is orthogonal to the line); then write $z$ in radial coordinates as  $(r,\theta)$.
 Recall \cite{MR895588} the notion of a
 {\em Poisson line process}
 (PLP) of intensity $\eta > 0$,
which makes precise the notion of ``completely random" lines in the plane.
 Parametrizing lines by by their closest points
 $(r,\theta)$, this PLP
 has intensity $\eta$ with respect to Lebesgue measure
 on parameter space
 $(0,\infty) \times (0,2\pi)$.
 The PLP distribution is invariant under Euclidean transformations, and for
 a fixed set $A$
 \begin{equation}
 \Ex (
 \mbox{length of line segments intersecting }A)
 = \pi \eta \times  \mbox{area}(A).
 \label{PLP-a}
 \end{equation}
 (We write $\Ex$ for expectation and $\Pr$ for probability).
 The next result shows how the PLP is useful in constructing spatial networks.
See Figure 1, right side.
\begin{Lemma}
\label{med-large}
Let $n^{1/2}/\sigma_n$ be an integer. 
Construct a network as the superposition of the rectangular grid with cell side-length $\sigma_n$ and the 
Poisson line process of intensity $\eta$, intersected with the square $[0,n^{1/2}]^2$.
Let $v_i, v_j$ be vertices of the grid.
Then 
\[ \Ex ( \mbox{route-length $v_i$ to $v_j$})
\leq |v_i - v_j| 
+ C_2  \sfrac{1}{\eta} \log (\eta \sqrt{2n}) \]
for an absolute constant $C_2$.
\end{Lemma}
Lemma \ref{med-large} is proved in \cite{me116}, Lemma 11, and we will not repeat 
the argument here. 
(In essence, one analyzes the natural routing algorithm: move to a nearby line of the PLP, move along that line in the direction closer to the direction of the destination city, and when encountering another line of the PLP, switch to that line if its direction is closer to the destination city direction).
Using the PLP gives us {\em random} networks, but a typical realization will have costs and lengths of the same order as the expectations in our formulas.

\subsection{Construction of the networks}
\label{sec-tc}
We now describe how the ingredients above 
(hierarchical routing on the square lattice, 
the PLP) 
are used in a network construction.
Recall $\bx^n$ denotes the given configuration of $n$ cities. 
Take integers $\theta_n \uparrow \infty$ slowly and define
\[ \sigma_n = n^{1/2}/\theta_n . \]
Let $M_n$ be the integer such that 
\[ \sigma_n/2 < 2^{M_n} \leq \sigma_n . \]
Define
\[ s_n = \sigma_n/2^{M_n}, \quad \quad 
\mbox{ (so $1 \leq s_n < 2$)} . \] 
Construct a network $G(\bx^n)$ as follows.

(i) Take the {\em large-scale} network in Lemma \ref{med-large}, with 
$\eta_n = \theta_n n^{-1/2}$.
 This network contains 
\emph{large cells} of side-length $\sigma_n$.

(ii) Inside each large cell put a copy of the hierarchical routing lattice of section \ref{sec-hier}, with $M = M_n$, and scaled so that the basic \emph{small cell} of this lattice has side-length $s_n$.

(iii) Link each city $x \in \bx^n$ via a straight edge to the bottom left corner vertex $v(x)$ of its small cell.

Figure 1 illustrates (i) and (ii).
There is a natural way to define a route from $x_i$ to $x_j$ in this network.  
From $x_i$ take the link to $v(x_i)$, then 
follow the section \ref{sec-hier} routing scheme to
the lower left corner $V(x_i)$ of the large cell; 
navigate from $V(x_i)$ to $V(x_j)$ via the shortest route in the Lemma \ref{med-large} graph.

Note that in addition to the given $n$ cities, this network has several different kinds of junctions:
 the vertices of the grid, and places where lines of the PLP cross each other or cross the grid lines or cross the short stage (iii) links.  In our model there is no cost associated with creating a junction or with routes using junctions; the costs involve only link lengths and route lengths.  So the exact number of junctions is unimportant.

\subsection{Analysis of the networks}
Clearly
\[ \ell(x_i,x_j) \leq \ell(V(x_i),V(x_j)) + 2^{3/2} \sigma_n
\]
and so by Lemma \ref{med-large}
\[ \Ex \ell(x_i,x_j) \leq |x_i - x_j|  + 2^{3/2} \sigma_n 
+ C_2  \sfrac{1}{\eta_n} \log (\eta_n \sqrt{2n}) .\] 
From the definitions of $\sigma_n,\eta_n$ we see 
\[ \Ex (\ell(x_i,x_j) - |x_i - x_j|) = o(n^{1/2}) \] 
establishing the modestly efficient property.

To analyze costs, we treat stages (i)-(iii) separately, and check that each stage cost  is less than the bounds stated
in Theorem \ref{T1}.

{\em Stage (iii).}
There are $n$ links of the form $(x,v(x))$, each carrying flow volume 
$2(1-\frac{1}{n}) n^{-1/2}$, 
and each having length at most $s_n \sqrt{2}$, and so \begin{equation} \mbox{the total cost of stage (iii) links is
$O(n^{1 - \frac{\beta}{2}})$.
} \label{cost-small}
\end{equation}

{\em Stage (ii).}
Now let $\EE_m$ be the set of type-$m$ edges. 
The number of such edges is 
$ \# \EE_m 
= O(n 2^{-m}) $. 
Recall that H\"{o}lder's inequality shows that for any edge-set $\EE$
\[
\sum_{e \in \EE}  f^\beta(e) \leq (\# \EE)^{1-\beta} \left(\sum_{e \in \EE} f(e) \right)^\beta .
\]
Now
\[ \sum_{e \in \EE_m} f(e) 
= 2n^{-1/2}  \sum_{x \in \bx^n} \#\{\mbox{ type-$m$ edges in route $v(x)$ to $V(x)$}\} 
\leq 2^{m+2} n^{1/2} 
\] 
using Lemma \ref{LmM}.  Thus
\begin{equation}
\sum_{e \in \EE_m}  f^\beta(e) = 
O\left( (n 2^{-m})^{1-\beta} \ (2^{m} n^{1/2} )^\beta \right)
= O\left( n^{1 - \frac{\beta}{2}} 2^{m(2\beta -1)} \right) . \label{cost-med}
\end{equation}
Writing $\EE_{med}$ for 
all edges in the copies of the hierarchical routing lattice, we find after summing over $0 \leq m \leq M$
\begin{eqnarray*}
\sum_{e \in \EE_{med}}  f^\beta(e) 
&=& O \left(n^{1 - \frac{\beta}{2}} \right), \quad 0 < \beta < \sfrac{1}{2} \\
&=& O \left(n^{3/4} \log n \right), \quad  \beta = \sfrac{1}{2} \\
&=& O \left(n^{1 - \frac{\beta}{2}}2^{M(2\beta -1)}\right) = O \left( n^{\frac{1}{2}+\frac{\beta}{2}}\right), \quad  \sfrac{1}{2} < \beta < 1
\end{eqnarray*}
using $2^M < n^{1/2}$. 
Because edge-lengths here are $s_n < 2$, these are bounds for the costs associated with stage (ii).

{\em Stage (iii).}
Write $\EE_{large}$ for the set of links of the large-scale network, that is the large-scale grid and the PLP lines.  
Flow along the route from $V(x_i)$ to $V(x_j)$ contributes 
$n^{-3/2} \ell(V(x_i),V(x_j))$ to the ``flow $\times$ distance" measure, and so 
\[ \int_{\EE_{large}} f(e) de 
= n^{-3/2} \sum_i \sum_j \ell(V(x_i),V(x_j)) \]
where the left side denotes integrating along all links of the large-scale network.  
By the already-established modestly efficient property,
\[  \sum_i \sum_j \ell(V(x_i),V(x_j)) = (1 + o(1)) \sum_i \sum_j |x_i-x_j| = O(n^{5/2}) \]
and so 
\[ \int_{\EE_{large}} f(e) de 
= O(n). \]
The total length $L_n$ of $\EE_{large}$ is the sum of 
$O(n^{1/2} \theta_n)$ ($=$ contribution from large-scale grid)  and  $O(\eta_n n)$ 
($=$ contribution from the PLP, using (\ref{PLP-a})), 
and so $L_n = O(n^{1/2} \theta_n)$.
The integral form of H\"{o}lder's inequality now shows that the cost associated
with $\EE_{large}$ is :
\begin{equation}
 \int_{\EE_{large}} f^\beta(e) de \leq L_n^{1-\beta} \times  \left(\int_{\EE_{large}} f(e) de \right)^\beta
= O\left( \theta_n^{1-\beta} n^{(1+\beta)/2} \right) .
\label{cost-large}
\end{equation}
Examining the cost of each stage, we check that the modestly efficient network we have constructed has its cost bounded as stated in 
Theorem \ref{T1}.  
Moreover, if we eliminate the ``modestly efficient" constraint then we can eliminate Stage (iii) of the construction (take $\theta_n = 1$) and get the stated $O(n^{\alpha(\beta)})$ upper bound.

\subsection{The transition at $\beta = 1/2$}
\label{sec-explain}
To summarize, the costs associated with 
the constructed networks 
arising from short, medium and large-scale links are bounded by expressions 
(\ref{cost-small},\ref{cost-med},\ref{cost-large}) 
respectively.
By examining the exponents of $n$
we see that the transition at $\beta = \frac{1}{2}$ corresponds to the dominant cost contribution changing from short links to long links.  
The arguments we give below for the lower bound show 
this is a genuine effect (no alternate networks can do essentially better),
not an artifact of the particular networks contructed above.

\section{The lower bound}
In the settings of \cite{me116,me118} the lower bounds require some effort to prove, but in the present setting the proofs are short.

\subsection{The case $0 < \beta \leq 1/2$}
Consider first the case $0 < \beta \leq 1/2$.
Impose the condition: there exists some small $\delta > 0$ such that
\begin{eqnarray}
&&\mbox{for at least $\delta n$ of the cities of $\bx^n$,}\nonumber\\
&&\mbox{ the distance to the nearest neighbor is at least $\delta$.}
\label{LB1}
\end{eqnarray}
Consider a city $x \in \bx^n$ satisfying this condition, and consider the link-segments of an arbitrary connected network within distance $\delta /2$ from $x$.   Because flow of volume $2n^{-1/2}$ must enter or leave $x$, the cost associated with these link-segments 
(which by concavity of $f \to f^\beta$ is minimized when there is a single link-segment) 
is at least $\delta/2 \times (2n^{-1/2})^\beta$.  Summing over all (there are at least $\delta n$) such cities $x$,  
noting the link-segments are distinct as $x$ varies, the network cost is at least 
$\delta n \times \delta/2 \times (2n^{-1/2})^\beta = \Omega(n^{1 - \frac{\beta}{2}})$.

\subsection{The case $1/2 < \beta < 1$}
In the case $1/2 < \beta < 1$ we impose
the classical equidistribution property for the configuration $\bx^n = (x^n_i,  1 \leq i \leq n)$ rescaled back to the unit square:
\begin{eqnarray}
&&\mbox{ the empirical distribution of $\{n^{-1/2}x^n_i, 1 \leq i \leq n\}$ converges} \nonumber\\
&&\mbox{ in distribution to the uniform
 distribution on $[0,1]^2$}.
 \label{LB2}
 \end{eqnarray}
 Our standardization conventions imply that the total volume of flow through the network is $\Theta(n^{1/2})$ and so assertion (a) below is obvious.

\begin{Lemma}
\label{Ln2}
(a) In the Gilbert network $\Gil(\bx^n)$, the maximum edge-flow is bounded as 
\[ \max_e f(e) = O(n^{1/2})  . \]
 (b) For any modestly efficient network $(G(\bx^n))$ on configurations satisfying the equidistribution condition (\ref{LB2}), the maximum edge-flow is bounded as
\[ \max_e f(e) = o(n^{1/2})  . \]
\end{Lemma}
Granted this result, use the fact
\[ \sum_e \ell(e) f(e) \geq  
n^{-3/2} \sum_i \sum_j |x_i-x_j| 
= \Theta(n) \mbox{ by equidistribution }  \]
and the general inequality
\[ \cost (G(\bx^n)) = \sum_e \ell(e) f^\beta(e) 
\geq \frac{\sum_e \ell(e) f(e)}{(\max_e f(e))^{1-\beta}} \]
to deduce that $\cost (G(\bx^n))$ grows strictly faster than 
$n/n^{(1-\beta)/2} = n^{\alpha(\beta)}$ for any modestly efficient network, and no slower than order 
$n^{\alpha(\beta)}$ for the Gilbert network.

\vspace{0.1in}
\noindent
{\bf Proof of Lemma \ref{Ln2}(b).}
  We first quote an easy fact from geometry.
 \begin{Lemma}
 \label{Lgeom}
 Let $Z_1,Z_2$ be two independent uniform random points in the unit square 
 $[0,1]^2$.  
 There exists a constant $C$ such that
for all $x \in [0,1]^2$ and all $\delta > 0$ 
\[ \Pr (|Z_1 - x| + |Z_2 - x| \leq |Z_1 - Z_2| + \delta)
\leq C \delta^{1/2} . \]
\end{Lemma}
Now fix $\delta > 0$.
Write $X_1,X_2$ for two uniform random picks from the set $\bx^n$ of cities.
The modestly efficient assumption implies
\[ \Pr (\ell(X_1,X_2) \geq |X_1 - X_2| + \delta n^{1/2})
\to 0 \mbox{ as } n \to \infty . \]
Lemma \ref{Lgeom} and the equidistribution assumption (\ref{LB2}) imply
\[ \Pr (|X_1 - x| + |X_2 - x| \leq |X_1 - X_2| + \delta \quad \mbox{ for all }  x)
\leq C \delta^{1/2} + o(1). \]
In order for the route from $X_1$ to $X_2$ to pass through point $x$, one of the two inequalities above must hold, and so
\[ \sup_x 
\Pr(\mbox{ route $X_1$ to $X_2$ passes through $x$ }) 
\leq C \delta^{1/2} + o(1). \]
But $\delta$ is arbitrary, so this probability is $o(1)$,
and the flow volume is exactly $n^{1/2}$ times this probability.

\section{Associating speeds with links}
\label{sec-speed}
The main feature of our model -- that the cost of building a link is sublinear in link capacity -- is just one of many realistic features one might want to incorporate into a model. 
By focussing on route lengths, we have implicitly assumed that users travel at constant speed. 
A notable feature of real road or rail networks is that different links permit different speeds. 
In this section we state and briefly discuss a variant model in which links can be designed to permit 
different speeds.

Suppose a link with length $\ell$, nominal capacity $c_0$ and nominal speed $s_0$ costs
$\ell c_0^\beta s_0^\gamma$, 
for fixed $0 < \gamma < \infty$.  
On such a link, traffic moves with speed $s_0$ provided the flow-volume $f$ is at most $c_0$; for larger flow-volumes, congestion causes the speed to drop, reaching speed zero (jammed) at volume 
$\sigma c_0$ for a constant $\sigma$.  So $\sigma c_0$ is the maximum capacity.
Precisely,
\[ \mbox{ speed at flow-volume } f = s_0 G(f/c_0)  \]
where $G(u) = 1$ for $0 \leq u \leq 1$ 
and $G(u)$ decreases from $1$ to $0$ as $u$ increases from $1$ to $\sigma$.
Otherwise the model is the same as before: we are given 
a configuration of $n$ cities in the square of area $n$, 
and we are required to route flow of volume $n^{-3/2}$ between each source-destination pair.

For any network and feasible routing, 
define average speed as
\[
\overline{\mathrm{speed}}
= \frac{\ave_{i,j} |x_i - x_j|}{\ave_{i,j} t(x_i,x_j)} 
\]
where $t(x_i,x_j)$ is the time taken to travel from $x_i$ to $x_j$.
For this model, we ask
\begin{quote}
What is the minimum cost for a network on a given configuration $\bx^n$ of cities that allows 
 $\overline{\mathrm{speed}} = s$?
\end{quote}
The answer is that, under the regularity assumptions of Theorem \ref{T1} 
(which are needed only for lower bounds), and  ignoring $O(\log n)$ terms.
\begin{equation}
\mbox{minimum cost grows as order 
$s^\gamma n^{\alpha^*(\beta,\gamma)}$, where }
\end{equation}
\begin{eqnarray*}
\alpha^*(\beta,\gamma) &=& 1 - \sfrac{\beta}{2} - \sfrac{\gamma}{2}, \quad 0 < 2\beta + \gamma  \leq 1\\
&=& \sfrac{1+\beta}{2}, \quad 1 \leq 2\beta + \gamma .
\end{eqnarray*}
Let us briefly indicate how the previous analysis is adapted to this setting.
Because costs scale with design speed $s_0$ as $s_0^\gamma$, it is enough to consider the
case $\overline{\mathrm{speed}}  = 1$, and show that minimum cost grows as order $n^{\alpha^*(\beta,\gamma)}$. 
To construct a network, use the networks constructed previously and assign design speeds as follows.
For links of the large-scale network, which routes will use for a distance of order $n^{1/2}$, design speed of order $1$.  
For type $m$ edges in the hierarchical routing lattice, which routes will use for a distance of order $2^m$, 
design speed of order $2^m n^{-1/2} \log n$.  
For the local links of the form $(x,v(x))$, which routes will use for distance $O(1)$, design speed of order $n^{-1/2}$.  
This ensures the typical times $t(x_i,x_j)$ are of order $n^{1/2}$ as required.  
To calculate the cost, we simply combine the previous estimates (\ref{cost-small},\ref{cost-med},\ref{cost-large}) of costs of providing flow-volumes of different links with the costs of the design speeds stipulated above; 
the total cost is of order 
\[ 
\hspace*{-0.5in} 
n^{1 - \frac{\beta}{2}} \times n^{-\frac{\gamma}{2}} 
\ + \ 
\sum_{m=0}^M n^{1 - \frac{\beta}{2}} 2^{m(2\beta -1)} \times (2^m n^{-1/2} \log n)^\gamma 
\ + \ 
n^{(1 + \beta)/2} \times 1 
\]
and this works out to be of the form $n^{\alpha^*(\beta,\gamma)}$ stated.

 \end{document}